\documentclass[prd,twocolumn,showpacs,superscriptaddress,nofootinbib,floatfix,showkeys,10pt]{revtex4-1}
\usepackage{graphicx}
\usepackage{amsmath}
\usepackage{bm}
\usepackage{yhmath}
\usepackage{mathtools}
\usepackage{wasysym}
\usepackage[colorlinks,citecolor=violet,urlcolor=violet,linkcolor=blue]{hyperref}
\usepackage{color}
\usepackage{cases}
\usepackage{subfigure}
\usepackage{times}
\usepackage{dcolumn,booktabs,bm}
\usepackage{slashed}
\usepackage{amsfonts,amssymb,stmaryrd,latexsym,amsmath}
\usepackage{textcomp}
\usepackage{multirow}
\usepackage{cancel}
\usepackage{array}
\usepackage{orcidlink}
\usepackage{physics}

\def\SU3{{\text{SU(3)}_{\rm F}}}

\def \pcs4338{{P_{\psi s}^\Lambda(4338)^0}}

\allowdisplaybreaks

\begin{document}
	
	\title{\textcolor{violet}{Eddington-inspired Born-Infeld gravity: Constraints from the generalized parton distributions (GPDs)}}
	
	\author{The MMGPDs\footnote{Modern Multipurpose GPDs} Collaboration:\\
	        Muhammad Goharipour\,\orcidlink{0000-0002-3001-4011}}\email{muhammad.goharipour@ipm.ir}    
    \thanks{Corresponding author}
	\affiliation{School of Physics, Institute for Research in Fundamental Sciences (IPM), P.O. Box  19395-5531, Tehran, Iran}
	\affiliation{School of Particles and Accelerators, Institute for Research in Fundamental Sciences (IPM), P.O. Box 19395-5746, Tehran, Iran}
	
	\author{Anoushiravan Moradi\,\orcidlink{0009-0004-6893-5571}}\email{a.moradi1992@ut.ac.ir}
	\affiliation{Department of Physics, University of Tehran, North Karegar Avenue, Tehran 14395-547, Iran}
	
	\author{K. Azizi\,\orcidlink{0000-0003-3741-2167}}\email{kazem.azizi@ut.ac.ir} 
	\affiliation{Department of Physics, University of Tehran, North Karegar Avenue, Tehran 14395-547, Iran}
	\affiliation{Department of Physics, Dogus University, Dudullu-\"{U}mraniye, 34775 Istanbul, T\"urkiye}
	\affiliation{School of Particles and Accelerators, Institute for Research in Fundamental Sciences (IPM), P.O. Box 19395-5746, Tehran, Iran}

	\begin{abstract}
The Eddington-inspired Born-Infeld (EiBI) theory of gravity modifies general relativity in high-density regimes. It offers an alternative framework that avoids cosmological singularities and remodels gravitational dynamics within compact objects. An important feature of EiBI gravity is its additional parameter, $\kappa$, which governs deviations from standard gravitational behavior. In this study, we investigate constraints on $\kappa$ using the internal pressure distribution of the proton, derived from gravitational form factor (GFF) $ D(t) $ obtained through a QCD analysis of generalized parton distributions (GPDs). By comparing pressure profiles extracted from skewness-dependent GPDs with previous determinations based on deeply virtual Compton scattering (DVCS) data, we establish updated bounds on $\kappa$. Our results show that the choice of proton pressure model significantly impacts the constraints, with the strongest limits ($|\kappa| \leq 0.10\text{--}0.3\, \text{m}^5\, \text{kg}^{-1}\, \text{s}^{-2}$). We further demonstrate that constraints obtained based on the first and second moments of the pressure distribution yield competitive bounds compared to those derived from peak pressures or those derived from just the first moment. These findings highlight the importance of precise experimental and theoretical determinations of the proton's mechanical properties in testing alternative theories of gravity. The present study motivates future improvements in GPD reconstructions for stronger constraints on EiBI gravity and related modifications.
\end{abstract}

	
	\maketitle
	
	\thispagestyle{empty}
	
\textit{\textbf{\textcolor{violet}{Introduction}}}~~
Since the publication of \textit{general relativity} (GR), Albert Einstein's geometric theory of gravitation, in 1915, numerous alternative theories have been proposed to address its limitations or explore different formulations. These include the Brans-Dicke theory~\cite{Brans:1961sx,Dicke:1961gz}, $f(R)$ gravity~\cite{Buchdahl:1970ldb,Starobinsky:1980te,Sotiriou:2008rp,DeFelice:2010aj}, Modified Newtonian Dynamics (MOND)~\cite{Milgrom:1983ca,Bekenstein:2004ne,Famaey:2011kh}, massive gravity~\cite{Fierz:1939ix,deRham:2010kj,deRham:2014zqa}, Horndeski theory~\cite{Horndeski:1974wa,Gleyzes:2014dya,Langlois:2018dxi}, Loop Quantum Gravity (LQG)~\cite{Ashtekar:1986yd,Ashtekar:2004eh,Perez:2012wv,Dapor:2017rwv}, Causal Dynamical Triangulations (CDT)~\cite{Ambjorn:1998xu,Ambjorn:2012jv,Loll:2019rdj,Ambjorn:2019pkp}, emergent gravity~\cite{Verlinde:2010hp,Verlinde:2016toy}, and Born-Infeld-$f(R)$ gravity~\cite{Makarenko:2014lxa,Kibaroglu:2024ico} (see Refs.~\cite{Nojiri:2010wj,Clifton:2011jh,Joyce:2014kja,Koyama:2015vza,Berti:2015itd,Nojiri:2017ncd,CANTATA:2021asi} for reviews).
In addition, Ba\~nados and Ferreira proposed a modified theory of gravity known as \textit{Eddington-Inspired Born-Infeld} (EiBI) gravity~\cite{Banados:2010ix}, which draws inspiration from the Born-Infeld electromagnetic action and Eddington's purely affine formulation of gravity. This theory avoids cosmological singularities and modifies gravitational dynamics in high-density regimes, such as the interiors of compact astrophysical objects (e.g., black holes and neutron stars) and the early universe~\cite{Pani:2011mg,Pani:2012qb,Harko:2013wka,Avelino:2012ue,Avelino:2012ge,Delsate:2012ky,Scargill:2012kg,Sham:2013sya,Bouhmadi-Lopez:2014jfa,Sotani:2014lua,Wei:2014dka,Du:2014jka,Avelino:2015fve,Avelino:2016kkj,BeltranJimenez:2017doy,Albarran:2018mpg,JimenezCruz:2019dnn}. It is worth noting in this context that EiBI gravity also predicts deviations from GR at subatomic scales, including within protons~\cite{Avelino:2012qe,Avelino:2019esh}.

In EiBI theory of gravity, the effective gravitational pressure can lead to a stronger gravitational field inside matter compared to GR~\cite{Avelino:2012ge,Delsate:2012ky}, despite both theories being identical in vacuum. To be more precise, EiBI represents a class of modified gravity theories that can predict moderate deviations from GR within the matter for appropriate values of
the EiBI's additional parameter, $\kappa$. The theory naturally reduces to GR in the limit $ \kappa \rightarrow 0 $. In Ref.~\cite{Avelino:2019esh}, it has been demonstrated that constraints on $\kappa$ can be derived from the pressure distribution of quarks within the proton, as measured via deeply virtual Compton scattering (DVCS) data~\cite{Burkert:2018bqq}. The resulting bound, $|\kappa| \lesssim c^4 |p_{\text{peak}}|^{-1} \sim 10^{-1}\, \text{m}^5\, \text{kg}^{-1}\, \text{s}^{-2}$ (where $ c $ is the speed of light and $ p_{\text{peak}} $ is the average peak pressure near the center of the proton), is nearly an order of magnitude stronger than limits inferred from neutron star core densities~\cite{Pani:2011mg} (and six order of magnitude stronger than limits inferred using solar physics~\cite{Casanellas:2011kf}), though weaker than those from other studies~\cite{Avelino:2012ge,Avelino:2012qe,Latorre:2017uve,Delhom:2019wir,BeltranJimenez:2021oaq}. While the connection between the internal gravitational field strength and proton pressure distributions is interesting, further investigation is needed to assess how different theoretical or experimental determinations of the proton's pressure profile might influence the constraints on $\kappa$.

The mechanical properties of the nucleon are encoded in the energy-momentum tensor (EMT) of quantum chromodynamics (QCD) through the gravitational form factors (GFFs). Among these, the $D$-form factor carries important information about the internal stress distribution, pressure, and shear forces within the nucleon~\cite{Polyakov:2018zvc,Burkert:2023wzr,Goharipour:2025lep,Dehghan:2025ncw}. This form factor can be extracted either from a QCD analysis of deeply virtual Compton scattering (DVCS) data~\cite{Burkert:2018bqq,Kumericki:2019ddg,Goharipour:2025lep} or through purely theoretical approaches such as light-cone QCD sum rules (LCSR)~\cite{Azizi:2019ytx,Dehghan:2025ncw} (for a brief review of different approaches, see Ref.~\cite{Goharipour:2025lep}). It is now well established that the $D$-form factor corresponds to the \textit{D-term} in the framework of generalized parton distributions (GPDs)~\cite{Belitsky:2005qn}. The $D$-term is a contribution to the unpolarized GPDs and governs their asymptotic behavior in the limit of the renormalization scale $\mu \rightarrow \infty$. GPDs are nonperturbative objects that offer a powerful framework for probing the three-dimensional (3D) structure of hadrons~\cite{Diehl:2003ny}. Unlike ordinary parton distribution functions (PDFs), which depend only on the longitudinal momentum fraction $x$ and the factorization scale $\mu$, GPDs incorporate two additional kinematic variables: $t$ (the negative momentum transverse squared) and $\xi$ (the skewness parameter representing the longitudinal momentum transverse). It should be noted that GPDs reduce to PDFs in the forward limit, where both $t$ and $\xi$ vanish.

In the MMGPDs Collaboration study~\cite{Goharipour:2025lep}, it was shown that the pressure distribution derived from the form factor $ D(t) $ obtained through a QCD analysis of skewness-dependent GPDs exhibits spatial variations in both magnitude and radial dependence compared to the pressure distribution reported in Ref.~\cite{Burkert:2018bqq}. The former analysis was based on Compton form factor (CFF) data, employing skewness-dependent GPDs constructed via the double-distribution (DD) representation~\cite{Radyushkin:1998es,Polyakov:1999gs}, with zero-skewness GPDs constrained by elastic scattering data~\cite{Hashamipour:2022noy}. In contrast, the latter study relied directly on DVCS data and a parameterization of Compton form factors (CFFs). Given these differences, it is interesting to explore how the revised proton pressure distribution might impact the constraint $|\kappa| \lesssim 10^{-1}\, \text{m}^5\, \text{kg}^{-1}\, \text{s}^{-2}$ derived in Ref.~\cite{Avelino:2019esh}.

In the present study, we first provide a concise review of the EiBI theory of gravity, with particular emphasis on its characterization of gravitational pressure within matter. Next, we outline the methodology for determining the proton's internal pressure distribution using the GFF $D(t)$, derived from a QCD analysis of skewness-dependent GPDs. Subsequently, we establish new constraints on the EiBI parameter $\kappa$ by analyzing different proton pressure profiles, comparing our results both internally and with the existing bound from Ref.~\cite{Avelino:2019esh}. Finally, we summarize our findings and present concluding remarks.\\
	
	
\textbf{\textit{\textcolor{violet}{EiBI theory of gravity and gravitational pressure}}}~~
This section provides a concise overview of the essential characteristics of EiBI gravity~\cite{Deser:1998rj,Vollick:2003qp,Vollick:2005gc,Vollick:2006qd,Banados:2010ix}, with special emphasis on the notion of effective gravitational pressure. In contrast to standard theories, EiBI provides a novel description of gravity under high-energy conditions and at the sub-femtometer scale~\cite{Kurbah:2024sdz}. To be more precise, EiBI maintains asymptotic agreement with GR while circumvents geodesic singularities even at the classical level~\cite{Deser:1998rj}. EiBI gravity is formulated through the action given by~\cite{Avelino:2019esh,Banados:2010ix}:
\begin{multline}
	S_{{\rm EiBI}} = \frac{1}{\kappa} \int d^4x \left[ 
	\sqrt{-\left|g_{\mu\nu} + \kappa \mathcal{R}_{\mu\nu}(\Gamma)\right|} \right. \\
	\left. - \lambda \sqrt{-|g_{\mu\nu}|} \right] + S_M(g, \Phi) ,\ \label{action}
\end{multline}
where \( S_M(g, \Phi) \) represents the matter action and \( \Phi \) refers to a general matter field. The parameter \( \lambda \) is dimensionless and related to the cosmological constant via \( \Lambda = (\lambda - 1)/\kappa \), while \( \kappa \) is an additional EiBI parameter with dimensions of length squared. Note that throughout this paper, we work in natural units where \( 8\pi G = c =1 \), except when stated otherwise. EiBI gravity is formulated in the Palatini approach~\cite{Olmo:2011uz}, wherein the metric tensor and the affine connection are regarded as independent dynamical variables. The equations of motion can be derived by performing independent variations of action~(\ref{action}) with respect to the connection \( \Gamma^\rho_{~\mu\nu} \) and the metric \( g_{\mu\nu} \). This yields, respectively, 
\begin{align}
	\sqrt{\frac{|q|}{|g|}}\, q^{\mu\nu} &= g^{\mu\nu} - \kappa T^{\mu\nu}, \label{eom1}
\end{align}
and
\begin{align}
	g_{\mu\nu} &= q_{\mu\nu} - \kappa \mathcal{R}_{\mu\nu}, \label{eom2}
\end{align}
where \( q_{\mu\nu} \) is the apparent metric used to construct the affine connection \( \Gamma^\rho_{~\mu\nu} \), and \( q^{\mu\nu} \) denotes its matrix inverse. Here,  \(g = \det(g_{\mu\nu})\) and \(q = \det(q_{\mu\nu})\) represent the determinants of the physical and apparent metrics, respectively. 
Note that since our focus is on a flat background, we set \( \lambda = 1 \) without loss of generality.

Combining Eqs.~(\ref{eom1}) and~(\ref{eom2}) one obtains the modified Einstein field equations:
\begin{equation}
	{{\mathcal G}^\mu}_\nu \equiv {{{\mathcal R}}^\mu}_\nu -\frac{1}{2} {\mathcal R} {\delta^\mu}_\nu  ={{\mathcal T}^{\mu}}_{\nu}\,. \label{fieldequation}
\end{equation}
In this expression, ${{\mathcal G}^\mu}_\nu$ denotes the components of the apparent Einstein tensor—analogous to the standard Einstein tensor in GR but constructed from the apparent metric—while ${{\mathcal T}^\mu}_\nu$ represents the components of the EMT, and ${\delta^\mu}_\nu$ is the Kronecker delta.
The Ricci tensor \( \mathcal{R}_{\mu\nu}(\Gamma) \) is constructed from the connection \( \Gamma \), and we have \( |{\cal G}_{\mu\nu}| = \det({\cal G}_{\mu\nu}) \). In the Palatini formalism, the connection components are constructed from apparent metric \( q_{\mu\nu} \) as follows:
\begin{equation}
	\Gamma^\rho_{~\mu\nu} = \frac{1}{2} q^{\rho\sigma} \left( \partial_\mu q_{\nu\sigma} + \partial_\nu q_{\mu\sigma} - \partial_\sigma q_{\mu\nu} \right). \label{Gamma_q}
\end{equation}
This condition ensures that the Ricci tensor \( \mathcal{R}_{\mu\nu}(\Gamma) \), appearing in the EiBI action, is constructed from the connection \( \Gamma^\rho_{~\mu\nu} \) and hence from the apparent metric \( q_{\mu\nu} \), rather than directly from the physical metric \( g_{\mu\nu} \).

In the context of EiBI gravity, the Poisson equation is modified to incorporate a correction term dependent on the matter distribution. The equation takes the form~\cite{Avelino:2019esh}:
\begin{equation}
\nabla^2 \phi = \frac{\rho}{2} + \frac{\kappa}{4} \nabla^2 \rho\,.
\end{equation}
where \( \phi(r) \) is the gravitational potential and \( \rho(r) \) is the energy density. The additional term \( \frac{\kappa}{4} \nabla^2 \rho(r) \) introduces a correction to the Newtonian potential. Note that depending on the energy density distribution function \( \rho(r) \),  the behaviour of  Laplacian acting on \( \rho(r) \),  and the sign of \( \kappa \),  the resulting  interaction can manifest as either attractive or repulsive at different distances from the center.

As demonstrated in Ref.~\cite{Delsate:2012ky}, the apparent EMT can be expressed in the form of a perfect fluid:
\begin{equation}
{{\mathcal T}^{\mu}}_{\nu}=(\rho_q+p_q)v^\mu v_\nu + p_q {\delta^{\mu}}_{\nu}\,,
\end{equation}
where \( v^\mu \) denotes the components of the apparent four-velocity of the fluid, satisfying the normalization condition \( v^\mu v^\nu q_{\mu \nu} = -1 \), and the corresponding apparent pressure and energy density are defined respectively as follows  \cite{Avelino:2019esh}:
\begin{eqnarray}
	p_q &=& \tau p + \mathcal{P}\,, \label{eq:pqrq}\\
	\rho_q &=& \tau \rho - \mathcal{P}\,, 
\end{eqnarray}
where the apparent pressure term $\mathcal{P}$ is given by
\begin{equation}
	\mathcal{P} = \frac{1}{\kappa}(\tau - 1) - \frac{1}{2} \tau (3p - \rho)\,. \label{eq:10}
\end{equation}
Here, the scalar quantity \(\tau\), which arises naturally during the derivation of the field equations, is defined as
\begin{equation}
	\tau \equiv \sqrt{\frac{g}{q}} = \left[ \det\left({\delta^\mu}_\nu - \kappa {T^\mu}_\nu\right) \right]^{-\frac{1}{2}}\,. \label{eq:11}
\end{equation}
In the case where the gravitational source is modeled as a perfect fluid, the scalar quantity \(\tau\) takes the form
\begin{equation}
	\tau = \left[ (1 + \kappa \rho)(1 - \kappa p)^3 \right]^{-\frac{1}{2}}\,, \label{eq:12}
\end{equation}
with \(\rho\) and \(p\) representing the physical energy density and pressure of the fluid, respectively.

Consequently, the apparent pressure $p_q$ can be interpreted as the sum of the physical pressure $p$ and an additional effective gravitational contribution $p_G$, defined as
\begin{equation}
	p_G \equiv p_q - p\,.
\end{equation}
Expanding Eqs.~(\ref{eq:pqrq}) to first order in $\kappa\rho$ and $\kappa p$, and substituting Eqs.~(\ref{eq:10}) and (\ref{eq:12}), the gravitational pressure correction $p_G$ becomes:
\begin{equation}
	p_G = \kappa p^2 + \frac{\kappa}{8} (\rho + p)^2\,. \label{eq:14}
\end{equation}
This expression reveals that the effective gravitational pressure $p_G$ is quadratic in the matter variables and becomes significant in regimes of high energy density or pressure. Note that, for typical astrophysical conditions where $\kappa\rho \ll 1$, these corrections are negligible. So they ensure consistency with GR in the low-energy limit. However, under extreme conditions such as those found in the early universe or the cores of compact stars, $p_G$ can contribute nontrivially to the gravitational dynamics, offering a possible way to avoid singularities or to modify collapse behavior.\\

	
\textbf{\textit{\textcolor{violet}{Pressure distribution inside the proton}}}~~ 
It is well established now that the pressure $p(r)$ and shear force $s(r)$  distributions inside the nucleon can be calculated using the GFF $ D(t) $ as follows~\cite{Polyakov:2018zvc,Burkert:2023wzr,Goharipour:2025lep,Dehghan:2025ncw,Shanahan:2018nnv,Lorce:2025oot}
\begin{align}
    p(r) &= p(r)=\frac{1}{6 m} \frac{1}{r^2}\frac{d}{dr} \left( r^2\frac{d}{dr}
 	{\widetilde{D}(r)} \right)\,, \label{eq:15} \\
    s(r) &= s(r)= -\frac{1}{4 m}\ r \frac{d}{dr} \left( \frac{1}{r} \frac{d}{dr}
	{\widetilde{D}(r)}\right) \,, \label{eq:16}
\end{align}
where $ \widetilde{D}(r) $ is the Fourier transform of $ D(t=-\bm{\Delta}^2) $ in the so-called Breit frame 
\begin{equation}
\label{eq:17}
{\widetilde{D}(r)=}
	\int {\frac{d^3\Delta}{(2\pi)^3}}\ e^{{-i} \bm{\Delta r}}\ D(-\bm{\Delta}^2)\,,
\end{equation}
and $ r $ is the radial distance from the nucleon center. These distributions  provide insight into the strong interaction dynamics that bind quarks
and gluons within hadrons and hence fundamental information about quark confinement and the mechanical stability of nucleons. Actually, due to confinement, the strong color field constantly generates quark-antiquark pairs and gluons inside the nucleon, whose interactions produce intense internal pressure. This dynamic sea contributes to the spatial distribution of forces that stabilize the nucleon against collapse or expansion. Consequently, the pressure distribution must satisfy the stability condition of the nucleon as
\begin{equation}
\label{eq:18}
    \int_0^\infty dr\, r^2  p(r)= 0\,, 
\end{equation}
which states that the total internal forces vanish when integrated over
the volume of the nucleon.

In Ref.~\cite{Goharipour:2025lep}, the MMGPDs Collaboration has recently investigated the proton's mechanical properties, including its mechanical and mass radii as well as its internal pressure and shear force distributions. To achieve this, they first extracted $ D(t) $ through a QCD analysis of available CFFs data, employing a model for skewness-dependent GPDs constructed using the DD representation. The DD representation generates the $ \xi $ dependence of GPDs by integration of a double-distribution function~\cite{Radyushkin:1998es,Polyakov:1999gs} and were successfully used for phenomenological analyses of the DVCS and DVMP data~\cite{Kroll:2012sm}. Then, they calculated the pressure distribution inside the proton for two values of parameter $ M $ that governs the $ t $-dependence of $ D(t) $ as follows 
\begin{equation}  
\label{eq:19}   
 D^Q(t)= \frac{4}{5} d_1^Q(0)\left(1 - \frac{t}{M^2} \right)^{-\alpha}\,,
\end{equation} 
where $ \alpha=3 $ and parameter $ d_1^Q(0) $ is determined from the fit to data. Note that their analysis just leads to the quark contribution of GFF $ D(t) $ and does not include the contribution from the gluon. 

\begin{figure}[t]
    \centering
\includegraphics[scale=0.7]{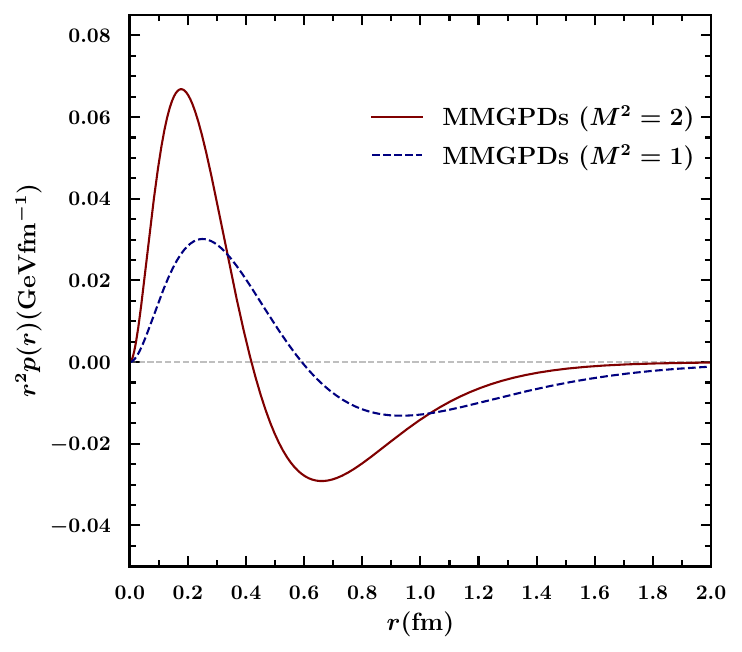}   
    \caption{Pressure distribution of the proton, $ r^2 p(r) $, as a function of the radial distance $  r$ from the centre of the proton obtained by the MMGPDs Collaboration~\cite{Goharipour:2025lep} from the GPD analysis for two values of the fit parameter $ M^2 $ in Eq.~(\ref{eq:19}).}.
\label{fig:pressure}
\end{figure}
Figure~\ref{fig:pressure} shows the results obtained in Ref.~\cite{Goharipour:2025lep} for the pressure distribution $ r^2 p(r) $  as a function of $ r $ for $ M^2=1,2 $ GeV$ ^2 $. As can be seen, the results obtained for 
$ M^2=2 $ GeV$ ^2 $ represents a remarkably strong repulsive pressure near the centre of the proton ($ r \lesssim 0.2 $ fm) and a binding pressure at greater distances. In the case of $ M^2=1 $ GeV$ ^2 $, smaller repulsive and binding pressures are obtained with  $ r \simeq 0.2 $ fm as the distance where the sign of pressure is changed. As mentioned before, these results are different from the corresponding one of Ref.~\cite{Burkert:2018bqq} that has been obtained by anlyzing the DVCS data and utilizing a parameterization of CFFs. 

\begin{figure}[t]
    \centering
\includegraphics[scale=0.7]{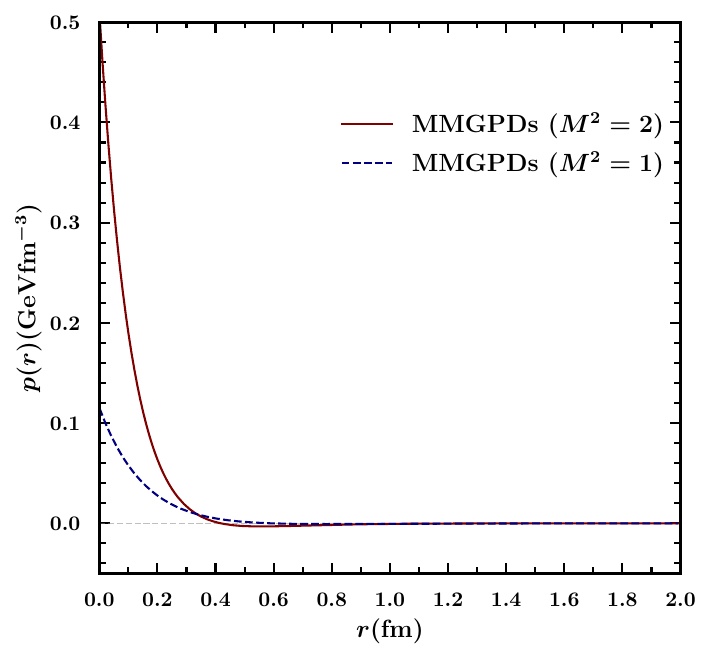}   
    \caption{Same as Fig.~\ref{fig:pressure} but for the pressure distribution $ p(r) $.}
\label{fig:Pr}
\end{figure}
Figure~\ref{fig:Pr} shows the same results as Fig.~\ref{fig:pressure} but for the pressure distribution $ p(r) $. As can be seen, the proton's pressure is not uniform; it peaks at the center and falls off rapidly with distance $ r $. Note that since the pressure diverges toward $ r=0 $, the ``peak" is often evaluated in a small but finite region near the center. According to this figure, the results obtained for $ M^2=1 $ and  $ M^2=2 $ GeV$ ^2 $ are significantly different near the center. Now the question is how these differences can affect the final result of Ref.~\cite{Avelino:2019esh} for parameter $ \kappa $. We address it in the following.\\


\textbf{\textit{\textcolor{violet}{Constraints on EiBI from proton interior pressure profile}}}~~ 
In this section, we investigate the potential constraints on EiBI gravity arising from measurements of the internal pressure profile of the proton. Special emphasis is placed on the role of quark pressure data. We also highlight how improved precision in measuring the pressure distribution within the proton~\cite{Goharipour:2025lep} could enhance these constraints. A foundational principle relevant to this discussion is the \textit{von Laue condition}, which states that the average pressure $\langle p \rangle$ inside stable compact objects with a negligible self-induced gravitational field must vanish. This condition has been explored in depth in Refs.~\cite{Polyakov:2018zvc,Avelino:2018qgt,Avelino:2018rsb}. Note that the average pressure for a spherically symmetric object of radius $ r_* $ and mass $ M_* $ can be written as 
\begin{equation}
	\langle p \rangle=\frac{\int_0^{r_*} p(r) r^2 dr}{\int_0^{r_*} r^2 dr}\,. \label{eq:avep}
\end{equation}

As shown, EiBI gravity behaves like GR but with an apparent metric governed by an apparent EMT. In such cases, the apparent energy density and pressure are comparable to or smaller than the corresponding physical quantities, unless the parameter $\tau$ in Eq.~(\ref{eq:12}) deviates significantly from unity. Consequently, the apparent metric within the proton in EiBI gravity closely resembles the Minkowski metric, and the von Laue condition remains valid if we replace the physical pressure $p$ with the total pressure as follows 
\begin{equation}
	p_T = p + p_G\,, \label{eq:p1}
\end{equation}
which leads to
\begin{equation}
	\langle p_T \rangle = 0\,, \label{eq:p2}
\end{equation}
and hence,
\begin{equation}
	|\langle p \rangle| = |\langle p_G \rangle|\,. \label{eq:p3}
\end{equation}
Therefore, any constraint on the average physical pressure inside the proton equivalently constrains the average effective gravitational pressure $p_G$. 

Let's start with a particular case. Considering the dominant energy condition, which requires $\rho \geq |p|$~\cite{Hawking:1973uf}, Eq.~(\ref{eq:11}) implies that the following constraint must be always satisfied in order to make $ \tau $ finite,
\begin{equation}
	|\kappa| \leq |p|^{-1}\,, \label{eq:const1}
\end{equation}
where we set $c = 1$ (note that $|p|^{-1}$ carries an implicit factor of $c^4$). 
Taking this constraint into account and using the proton's internal pressure profiles derived from the GPD analysis in Ref.~\cite{Goharipour:2025lep}, one can determine the upper limit of $|\kappa|$. 
If we take the maximum of the radial pressure profile $p(r)$ in Fig.~\ref{fig:Pr} ($ r\rightarrow 0 $ ), the resulting bounds on $|\kappa|$ for the two values of $ M^2 $ are:
\begin{align}
   |\kappa| \leq 0.44\, \text{m}^5\, \text{kg}^{-1}\, \text{s}^{-2}\,, ~~~~~~~ M^2=1 {\text{~GeV}}^2 \,, \label{eq:25} \\
   |\kappa| \leq 0.10\, \text{m}^5\, \text{kg}^{-1}\, \text{s}^{-2}\,, ~~~~~~~ M^2=2 {\text{~GeV}}^2 \,. \label{eq:26}
\end{align}
While if we use the the maximum of the radial pressure profile $r^2 p(r)$ in Fig.~\ref{fig:pressure} that occurs at $ r=0.25 $ and $ r=0.177 $ fm for $ M^2=1 $ and  $ M^2=2 $ GeV$ ^2 $, respectively, we obtain:
\begin{align}
   |\kappa| \leq 2.69\, \text{m}^5\, \text{kg}^{-1}\, \text{s}^{-2}\,, ~~~~~~~ M^2=1 {\text{~GeV}}^2 \,, \label{eq:27} \\
   |\kappa| \leq 0.61\, \text{m}^5\, \text{kg}^{-1}\, \text{s}^{-2}\,, ~~~~~~~ M^2=2 {\text{~GeV}}^2 \,. \label{eq:28}
\end{align}

Alternatively, if the average peak pressure is defined via Eq.~(\ref{eq:avep}), where $r_*$ marks the position of the peak in Fig.~\ref{fig:pressure}, the constraints become:  
\begin{align}
   |\kappa| \leq 1.56\, \text{m}^5\, \text{kg}^{-1}\, \text{s}^{-2}\,, ~~~~~~~ M^2=1 {\text{~GeV}}^2 \,, \label{eq:29} \\
   |\kappa| \leq 0.35\, \text{m}^5\, \text{kg}^{-1}\, \text{s}^{-2}\,, ~~~~~~~ M^2=2 {\text{~GeV}}^2 \,. \label{eq:30}
\end{align}
As can be seen, overall, the pressure distribution belonging to $ M^2=2 $ GeV$ ^2 $ leads to stronger limits than $ M^2=1 $ GeV$ ^2 $.

It is well established in Ref.~\cite{Avelino:2019esh} that the proton can be modeled as a spherically symmetric, spin-zero compact object rather than a perfect fluid. Considering the fact that only the diagonal components of the apparent EMT are relevant, the following relation is obtained for the apparent pressure
\begin{align}
	p_q &= \frac{{\mathcal{T}^i}_i}{3} = \frac{1}{\kappa}(\tau - 1) + \tau \left( \frac{{T^i}_i}{3} - \frac{{T^\mu}_\mu}{2} \right) \nonumber \\
	&= \frac{1}{\kappa}(\tau - 1) + \tau \frac{\rho - p}{2}\,, \label{eq:pqnew}
\end{align}
where $p \equiv {{{T}^{i}}_{i}}/{3}$ and ${T^{\mu}}_{\mu} = -\rho + 3p$, for any fluid at rest (not necessarily a perfect fluid). Using Eqs.~(\ref{eq:12}) and (\ref{eq:pqnew}) and substituting the components in spherical symmetry, Avelino obtained the following constraint~\cite{Avelino:2019esh}:
	\begin{equation}
		|p_G| \ge |\kappa| p^2\,. \label{eq:pgp2}
	\end{equation}
Through further analysis, a new constraint on the parameter \( \kappa \) can be obtained as follows
\begin{equation}
	|\kappa| \le \frac{|\langle p \rangle|}{\langle p^2 \rangle} = \frac{\xi}{\sqrt{\langle p^2 \rangle}}\,, \label{eq:kconst}
\end{equation}
	where the dimensionless parameter 
\begin{equation}
	\xi \equiv \frac{\langle p \rangle}{\sqrt{\langle p^2 \rangle}}\,,
\end{equation}
quantifies the degree of pressure anisotropy within the proton. It provides a measure of how gravitational interactions affect the internal pressure distribution and structure. As mentioned in Ref.~\cite{Avelino:2019esh}, Eq.~(\ref{eq:kconst}) provides a weaker constraint compared to Eq.~(\ref{eq:const1}), as it involves the measurement of the first and second moments of the physical pressure. We examined this statement using the pressure distributions obtained in Ref.~\cite{Goharipour:2025lep} and found the following bonds:
\begin{align}
   |\kappa| \leq 1.32\, \text{m}^5\, \text{kg}^{-1}\, \text{s}^{-2}\,, ~~~~~~~ M^2=1 {\text{~GeV}}^2 \,, \label{eq:35} \\
   |\kappa| \leq 0.30\, \text{m}^5\, \text{kg}^{-1}\, \text{s}^{-2}\,, ~~~~~~~ M^2=2 {\text{~GeV}}^2 \,. \label{eq:36}
\end{align}
Comparing the results of Eqs.~(\ref{eq:29}) and~(\ref{eq:30}) with the corresponding ones of Eqs.~(\ref{eq:35}) and~(\ref{eq:36}), one concludes that the constraint of Eq.~(\ref{eq:kconst}) leads to stronger limit on $ |\kappa| $ than the constraint of Eq.~(\ref{eq:const1}), but the difference is not remarkable. Overall, the results obtained here indicates that further experimental and phenomenological efforts are needed to achieve the precise reconstruction of the quark and gluon pressure distributions within the proton.
Actually, these improvements are expected to lead to more stringent bounds on modified gravity theories, including tighter constraints on the EiBI coupling parameter $\kappa$. Note that, while the mechanical properties of the proton such as the mechanical and mass radii as well as the pressure and shear force distributions  have been taken into account both theoretically and phenomenologically~\cite{Goharipour:2025yxm,Goharipour:2025lep,Dehghan:2025ncw}, the pure experimental determination of the first and second moments of the pressure distribution, which are essential for a more accurate and robust bound on $\kappa$, are not yet available. \\

	
\textit{\textbf{\textcolor{violet}{Conclusions}}}~~
In this work, we have explored the constraints on the EiBI gravity parameter $\kappa$ using the proton's internal pressure distribution.  To this aim, we have used the pressure distributions determined by the MMGPDs Collaboration~\cite{Goharipour:2025lep} through the QCD analysis of GPDs.
Our study reveals that the bounds on $\kappa$ are sensitive to the specific proton pressure model employed. The stronger limits are obtained for higher values of parameter $ M^2 $ ($M^2 = 2\, \text{GeV}^2$), which governs the $ t $-dependence of GFF $ D(t) $,  compared to lower ones ($M^2 = 1\, \text{GeV}^2$). The derived constraints, ranging from $|\kappa| \leq 0.10\, \text{m}^5\, \text{kg}^{-1}\, \text{s}^{-2}$ (for peak pressures) to $|\kappa| \leq 0.3\, \text{m}^5\, \text{kg}^{-1}\, \text{s}^{-2}$ (for averaged pressures), are competitive with existing bounds from neutron stars and astrophysical observations~\cite{Pani:2011mg}, though they remain several orders of magnitude weaker than those obtained from collider experiments~\cite{BeltranJimenez:2017doy,Latorre:2017uve}.

In particular, we found that constraints based on the moments of the pressure distribution, Eq.~(\ref{eq:kconst}), do not significantly tighten the bounds compared to those from peak pressure considerations, contrary to initial expectations. This suggests that further refinements in the experimental determination of proton pressure distributions, particularly the first and second moments, are critical for more precise tests of EiBI gravity. We conclude that future advances in GPD extractions, combined with higher-precision DVCS data and lattice QCD calculations, can provide important information on this issue and strengthen the constraints on $\kappa$.

Our results underscore the potential of subatomic physics, particularly the study of nucleon structure, for providing a novel testing ground for modified gravity theories. The present study highlights the importance of interdisciplinary approaches in probing the fundamental nature of gravity and matter as well as the need for further investigation on the gap between gravitational physics and QCD.\\


\textit{\textbf{\textcolor{violet}{Acknowledgements}}}~~
M.~Goharipour is thankful to the School of Particles and Accelerators and School of Physics, Institute for Research in Fundamental Sciences (IPM), for financial support provided for this research.
K. Azizi thanks Iran national science foundation (INSF) for the partial financial support provided under the elites Grant No. 4037888.	
\onecolumngrid
		
\twocolumngrid


\bibliographystyle{apsrev4-1}
\bibliography{article} 
	
\onecolumngrid

\end{document}